\documentclass[10pt,conference]{IEEEtran}

\usepackage{amsmath}
\usepackage{amssymb}
\usepackage{graphicx}
\usepackage{psfrag}
\usepackage{subfigure}
\usepackage{url}
\usepackage{cite}

\newcommand{\C}{\ensuremath{\mathcal{C}}}
\newcommand{\Ct}[1]{\ensuremath{C^{#1}_t}}
\newcommand{\Cr}[1]{\ensuremath{C^{#1}_r}}
\newcommand{\Rt}[1]{\ensuremath{R^{#1}_t}}
\newcommand{\Rr}[1]{\ensuremath{R^{#1}_r}}
\newcommand{\Rpr}[1]{\ensuremath{R'_r}}
\newcommand{\Cn}{\ensuremath{C_n}}
\newcommand{\ns}{\!}

\newcommand{\E}{\ensuremath{\mathrm{E}}}

\providecommand{\abs}[1]{\lvert#1\rvert}

\begin{document}

\title{Capacity Gain from Transmitter and\\Receiver Cooperation}

\author{\authorblockN{Chris T. K. Ng and Andrea J. Goldsmith}
\authorblockA{Dept.\ of Electrical Engineering\\
Stanford University, Stanford, CA 94305\\
Email: \{ngctk, andrea\}@wsl.stanford.edu}
}

\maketitle

\begin{abstract}
  Capacity gain from transmitter and receiver cooperation are compared
  in a relay network where the cooperating nodes are close together.
  When all nodes have equal average transmit power along with full
  channel state information (CSI), it is proved that transmitter
  cooperation outperforms receiver cooperation, whereas the opposite
  is true when power is optimally allocated among the nodes but only
  receiver phase CSI is available. In addition, when the nodes have
  equal average power with receiver phase CSI only, cooperation is
  shown to offer no capacity improvement over a non-cooperative scheme
  with the same average network power. When the system is under
  optimal power allocation with full CSI, the decode-and-forward
  transmitter cooperation rate is close to its cut-set capacity upper
  bound, and outperforms compress-and-forward receiver cooperation.
  Moreover, it is shown that full CSI is essential in transmitter
  cooperation, while optimal power allocation is essential in receiver
  cooperation.
\end{abstract}

\section{Introduction}
\label{sec:intro}

In ad-hoc wireless networks, cooperation among nodes can be exploited
to improve system performance, and the benefits of transmitter and
receiver cooperation have been recently investigated by several
authors. The idea of cooperative diversity was pioneered in
\cite{sendonaris03:coop1, sendonaris03:coop2}, where the transmitters
cooperate by repeating detected symbols of other transmitters. In
\cite{hunter02:coop_coding} the transmitters forward parity bits of
the detected symbols, instead of the entire message, to achieve
cooperation diversity. Cooperative diversity and outage behavior was
studied in \cite{laneman04:coop_diver}. Multiple-antenna systems and
cooperative ad-hoc networks were compared in \cite{jindal04:cap_coop,
  ng04:txcoop_dcp_relay}.  Information-theoretic achievable rate
regions and bounds were derived in \cite{host-madsen04:coop_bounds,
  khojastepour04:coop_relay, host-madsen04:power_relay,
  host-madsen04:rx_coop, host-madsen03:coop_rate} for channels with
transmitter and/or receiver cooperation. In
\cite{kramer04:coop_cap_relay} cooperative strategies for relay
networks were presented.

In this paper, we consider the case in which a relay can be deployed
either near the transmitter, or near the receiver. Hence unlike
previous works where the channel was assumed given, we treat the
placement of the relay, and thus the resulting channel, as a design
parameter. Capacity improvement from cooperation is considered under
system models of full or partial channel state information (CSI), with
optimal or equal power allocation.

\section{System Model}
\label{sec:sysmodel}

\begin{figure}
  \centerline{
    \subfigure[Transmitter cooperation]{
      \psfrag{e1}[][]{$e^{j\theta_1}$}
      \psfrag{e2}[][]{$\sqrt{g}e^{j\theta_2}$}
      \psfrag{e3}[][]{$e^{j\theta_3}$}
      \includegraphics{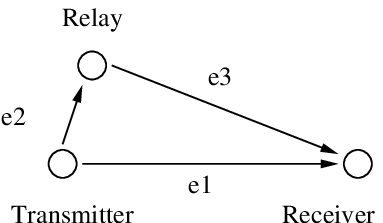}
      \label{fig:txcoop}
    }
    \hfil
    \subfigure[Receiver cooperation]{
      \psfrag{e1}[][]{$e^{j\theta_1}$}
      \psfrag{e2}[][]{$e^{j\theta_2}$}
      \psfrag{e3}[][]{$\sqrt{g}e^{j\theta_3}$}
      \includegraphics{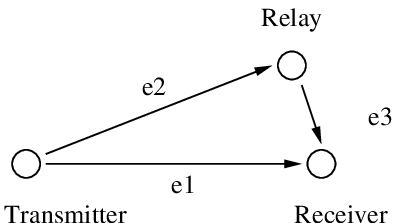}
      \label{fig:rxcoop}
    }
  }
  \caption{Cooperation system model.}
  \label{fig:coop}
\end{figure}

Consider a discrete-time additive white Gaussian noise (AWGN) wireless
channel. To exploit cooperation, a relay can be deployed either close
to the transmitter to form a transmitter cluster, or close to the
receiver to form a receiver cluster, as illustrated in
Fig.~\ref{fig:coop}. In the transmitter cluster configuration, suppose
the channel magnitude between the cluster and the receiver is
normalized to unity, while within the cluster it is denoted by
$\sqrt{g}$. The transmitter cooperation relay network in
Fig.~\ref{fig:txcoop} is then described by
\begin{align}
  y_1 & = \sqrt{g}e^{j\theta_2}x + n_1, & 
  y & = e^{j\theta_1}x + e^{j\theta_3}x_1 + n,
\end{align}
where $x,y,n,x_1,y_1,n_1\in\mathbb{C}$, $g\geq 0$, and $\theta_1,
\theta_2, \theta_3 \in [0,2\pi]$: $x$ is the signal sent by the
transmitter, $y$ is the signal received by the receiver, $y_1, x_1$
are the received and transmitted signals of the relay, respectively,
and $n, n_1$ are independent zero-mean unit-variance complex Gaussian
random variables. Similarly, the receiver cooperation relay network in
Fig.~\ref{fig:rxcoop} is given by
\begin{align}
  y_1 & = e^{j\theta_2}x + n_1, &
  y & = e^{j\theta_1}x + \sqrt{g}e^{j\theta_3}x_1 + n.
\end{align}
The output of the relay depends causally on its past inputs, and there
is an average network power constraint on the system:
$\E\bigl[\abs{x}^2 + \abs{x_1}^2\bigr] \leq P$, where the expectation
is taken over repeated channel uses.

We compare the rate achieved by transmitter cooperation versus that by
receiver cooperation under different operational environments. We
consider two models of CSI: i) every node has full CSI; ii) only
receiver phase CSI is available (i.e., the relay knows $\theta_2$, the
receiver knows $\theta_1, \theta_3$, and $g$ is assumed to be known to
all). In addition, we also consider two models of power allocation: i)
power is optimally allocated between the transmitter and the relay,
i.e., $\E\bigl[\abs{x}^2\bigr] \leq \alpha P$,
$\E\bigl[\abs{x_1}^2\bigr]\leq (1-\alpha) P$, where $\alpha \in [0,1]$
is a parameter to be optimized; ii) the network is homogeneous and all
nodes have equal average power constraints, i.e.,
$\E\bigl[\abs{x}^2\bigr]$ = $\E\bigl[\abs{x_1}^2\bigr]$ = $P/2$.
Power allocation in an AWGN relay network with arbitrary channel gains
was treated in \cite{host-madsen04:power_relay}; in this paper we only
consider the case when the cooperating nodes form a cluster.
Combining the different considerations of CSI and power allocation
models, Table~\ref{tab:cases} enumerates the four cases under which
the benefits of transmitter and receiver cooperation are investigated
in the next section.

\begin{table}
  \renewcommand{\arraystretch}{1.2}
  \centering
  \begin{tabular}{|c|l|}
    \hline
    \emph{Case} & \emph{Description}\\
    \hline\hline
    Case~1 & Optimal power allocation with full CSI\\
    \hline
    Case~2 & Equal power allocation with full CSI\\
    \hline
    Case~3 & Optimal power allocation with receiver phase CSI\\
    \hline
    Case~4 & Equal power allocation with receiver phase CSI\\
    \hline
  \end{tabular}
  \caption{Cooperation under different operational environments} 
  \label{tab:cases}
\end{table}

\section{Cooperation Strategies}
\label{sec:coopstrat}

The three-terminal networks shown in Fig.~\ref{fig:coop} are relay
channels \cite{meulen71:3t_comm_ch, cover79:cap_relay}, and their
capacity is not known in general. The cut-set bound described in
\cite{cover79:cap_relay, cover91:eoit} provides a capacity upper
bound. Achievable rates obtained by two coding strategies were also
given in \cite{cover79:cap_relay}. The first strategy
\cite[Thm.~1]{cover79:cap_relay} has become known as (along with other
slightly varied nomenclature) ``decode-and-forward''
\cite{laneman04:coop_diver, kramer04:coop_cap_relay,
  host-madsen04:coop_bounds}, and the second one
\cite[Thm.~6]{cover79:cap_relay} ``compress-and-forward''
\cite{kramer04:coop_cap_relay, host-madsen04:power_relay,
  khojastepour04:coop_relay}. In particular, it was shown in
\cite{kramer04:coop_cap_relay} that decode-and-forward approaches
capacity (and achieves capacity under certain conditions) when the
relay is near the transmitter, whereas compress-and-forward is close
to optimum when the relay is near the receiver. Therefore, in our
analysis decode-and-forward is used in transmitter cooperation, while
compress-and-forward is used in receiver cooperation.

Notations for the upper bounds and achievable rates are summarized in
Table~\ref{tab:rate_notations}. A superscript is used, when
applicable, to denote which case listed in Table~\ref{tab:cases} is
under consideration; e.g., $\Ct{1}$ corresponds to the transmitter
cut-set bound in Case~1. For comparison, $\Cn$ represents the
non-cooperative channel capacity when the relay is not available and
the transmitter has average power $P$; hence $\Cn=\C(1)$, where $\C(x)
\triangleq \log_2(1+xP)$.

\begin{table}
  \renewcommand{\arraystretch}{1.2}
  \centering
  \begin{tabular}{|c|l|}
    \hline
    \emph{Notation} & \emph{Description}\\
    \hline\hline
    $\Ct{}$ & Transmitter cooperation cut-set bound\\
    \hline
    $\Rt{}$ & Decode-and-forward transmitter cooperation rate\\
    \hline
    $\Cr{}$ & Receiver cooperation cut-set bound\\
    \hline
    $\Rr{}$ & Compress-and-forward receiver cooperation rate\\
    \hline
    $\Cn$ & Non-cooperative channel capacity\\
    \hline
  \end{tabular}
  \caption{Notations for the upper bounds and achievable rates}
  \label{tab:rate_notations}
\end{table}

Suppose that the transmitter is operating under an average power
constraint $\alpha P$, $0\leq\alpha\leq 1$, and the relay under
constraint $(1-\alpha)P$. Then for the transmitter cooperation
configuration depicted in Fig.~\ref{fig:txcoop}, the cut-set bound is
\begin{align}
  \label{eq:Ct}
  \begin{split}
    \Ct{} = \max_{\substack{0\leq\rho\leq 1}}
    \min\Bigl\{\C\Bigl(\alpha(g+1)(1-\rho^2)\Bigr),\Bigr.\\
    \Bigl.\C\Bigl(1+2\rho\sqrt{\alpha(1-\alpha)}\Bigr)\Bigr\},
  \end{split}
\end{align}
where $\rho$ represents the correlation between the transmitted
signals of the transmitter and the relay. With optimal power
allocation in Case~1 and Case~3, $\alpha$ is to be further optimized,
whereas $\alpha=1/2$ in Case~2 and Case~4 under equal power
allocation.

In the decode-and-forward transmitter cooperation strategy,
transmission is done in blocks: the relay first fully decodes the
transmitter's message in one block, then in the ensuing block the
relay and the transmitter cooperatively send the message to the
receiver. The following rate can be achieved:
\begin{align}
  \label{eq:Rt}
  \begin{split}
    \Rt{} = \max_{\substack{0\leq\rho\leq 1}}
    \min\Bigl\{\C\Bigl(\alpha g(1-\rho^2)\Bigr),
    \C\Bigl(1+2\rho\sqrt{\alpha(1-\alpha)}\Bigr)\Bigr\},
  \end{split}
\end{align}
where $\rho$ and $\alpha$ carry similar interpretations as described
above in (\ref{eq:Ct}). Note that $\Rt{}\bigr\rvert_{g} =
\Ct{}\bigr\rvert_{g-1}$ for $g\geq 1$, which can be used to aid the
calculation of $\Rt{}$ in the subsequent sections.

For the receiver cooperation configuration shown in
Fig.~\ref{fig:rxcoop}, the cut-set bound is
\begin{align}
  \label{eq:Cr}
  \begin{split}
    \Cr{} = & \max_{\substack{0\leq\rho\leq 1}}
    \min\Bigl\{\C\Bigl(2\alpha(1-\rho^2)\Bigr),\Bigr.\\
    \Bigl.&\C\Bigl(\alpha+(1-\alpha)g
    +2\rho\sqrt{\alpha(1-\alpha)g}\Bigr)\Bigr\}.
  \end{split}
\end{align}

In the compress-and-forward receiver cooperation strategy, the relay
sends a compressed version of its observed signal to the receiver. The
compression is realized using Wyner-Ziv source coding
\cite{wyner76:rate_dist_side_info}, which exploits the correlation
between the received signal of the relay and that of the receiver. The
following rate is achievable:
\begin{align}
  \label{eq:Rr}
  \Rr{} = \C\Bigl(\textstyle\frac{\alpha(1-\alpha)g}
    {(1-\alpha)g+2\alpha+1/P}+\alpha\Bigr).
\end{align}
Likewise, in (\ref{eq:Cr}) and (\ref{eq:Rr}) $\alpha$ needs to be
optimized in Case~1 and Case~3, and $\alpha=1/2$ in Case~2 and Case~4.

% fixthis: add transition

\subsection*{Case 1: Optimal power allocation with full CSI}
\label{sec:case1}

Consider the transmitter cooperation cut-set bound in (\ref{eq:Ct}).
Recognizing the first term inside $\min\{\cdot\}$ is a decreasing
function of $\rho$, while the second one is an increasing one, the
optimal $\rho^*$ can be found by equating the two terms (or maximizing
the lesser term if they do not equate). Next the optimal $\alpha^*$
can be calculated by setting its derivative to zero. The other upper
bounds and achievable rates, unless otherwise noted, can be optimized
using similar techniques; thus in the following sections they will be
stated without repeating the analogous arguments.

The transmitter cooperation cut-set bound is found to be
\begin{align}
  \label{eq:Ct1}
  \Ct{1} = \C\bigl(\textstyle\frac{2(g+1)}{g+2}\bigr),
\end{align}
with $\rho^* = \sqrt{g/(g+4)}$, $\alpha^* = (g+4)/(2g+4)$. The
decode-and-forward transmitter cooperation rate is
\begin{align}
  \label{eq:Rt1}
  \Rt{1} = \Bigl\{\textstyle 
  \C\bigl(\frac{2g}{g+1}\bigr) \text{ if } g\geq1, 
  \quad \C(g) \text{ if } g<1 \Bigr.,
\end{align}
with $\rho^*=\sqrt{(g-1)/(g+3)}$, $\alpha^*=(g+3)/(2g+2)$ if $g\geq1$,
and $\rho^*=0$, $\alpha^*=1$ otherwise.  It can be observed that the
transmitter cooperation rate $\Rt{1}$ in (\ref{eq:Rt1}) is close to
its upper bound $\Ct{1}$ in (\ref{eq:Ct1}) when $g\gg1$.

For receiver cooperation, the cut-set bound is given by
\begin{align}
  \label{eq:Cr1}
  \Cr{1} = \C\bigl(\textstyle\frac{2(g+1)}{g+2}\bigr),
\end{align}
with $\rho^* = 1/\sqrt{g^2+2g+2}$, $\alpha^*=(g^2+2g+2)/(g^2+3g+2)$.

The expression of the optimal value $\alpha^*$ for the
compress-and-forward receiver cooperation rate in (\ref{eq:Rr}) is
complicated, and does not facilitate straightforward comparison of
$\Rr{1}$ with the other upper bounds and achievable rates. A simpler
upper bound to $\Rr{1}$, however, can be obtained by omitting the term
$1/P$ in the denominator in (\ref{eq:Rr}) as follows:
\begin{align}
  \label{eq:Rr1}
  \Rr{1} & = {\max_{0\leq\alpha\leq1}}
  \C\Bigl(\textstyle
    \frac{\alpha(1-\alpha)g}{(1-\alpha)g+2\alpha+1/P}+\alpha\Bigr)\\
  \label{eq:Rr1_def}
  & < {\max_{0\leq\alpha\leq1}}\C\Bigl(\textstyle
  \frac{\alpha(1-\alpha)g}{(1-\alpha)g+2\alpha}+\alpha\Bigr) 
  \triangleq \Rpr\ .
\end{align}
Since the term $(1-\alpha)g+2\alpha$ in the denominator in
(\ref{eq:Rr1}) ranges between 2 and $g$, the upper bound $\Rpr\ $ in
(\ref{eq:Rr1_def}) is tight when $g>2$ and $P\gg1$. Specifically, for
$g>2$, the receiver cooperation rate upper bound is found to be
\begin{align}
  \label{eq:Rpr}
  \Rpr\ = \C\Bigl(\textstyle
  \frac{2g(\sqrt{g-1}-1)(g-1-\sqrt{g-1})}{\sqrt{g-1}(g-2)^2}\Bigr),
\end{align}
with the upper bound's optimal
$\alpha^*=\frac{g(g-1-\sqrt{g-1})}{g^2-3g+2}$.

Note that the transmitter and receiver cut-set bounds $\Ct{1}$ and
$\Cr{1}$ are identical. However, for $\infty>g>1$, it can be shown
that the decode-and-forward transmitter cooperation rate $\Rt{1}$
outperforms the compress-and-forward receiver cooperation upper bound
$\Rpr\ $. Moreover, the decode-and-forward rate is close to the
cut-set bounds when $g\geq2$; therefore, transmitter cooperation is
the preferable strategy when the system is under optimal power
allocation with full CSI\@.

\begin{figure}
  \centering
  \includegraphics[width=8cm]{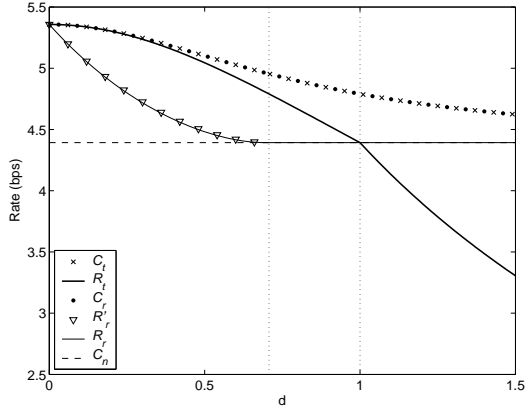}
  \caption{Cut-set bounds and achievable rates in Case~1.}
  \label{fig:rates_1}
\end{figure}

Numerical examples of the upper bounds and achievable rates are shown
in Fig.~\ref{fig:rates_1}. In all plots of the numerical results, we
assume the channel has unit bandwidth, the system has an average
network power constraint $P$ = 20, and $d$ is the distance between the
relay and its cooperating node. We assume a path-loss power
attenuation exponent of 2, and hence $g=1/d^2$. The vertical dotted
lines mark $d=1/\sqrt{2}$ and $d=1$, which correspond to $g=2$ and
$g=1$, respectively. We are interested in capacity improvement when
the cooperating nodes are close together, and $d<1/\sqrt{2}$ (or
$g>2$) is the region of our main focus.
 
\subsection*{Case 2: Equal power allocation with full CSI}
\label{sec:case2}

With equal power allocation, both the transmitter and the relay are
under an average power constraint of $P/2$, and so $\alpha$ is set to
$1/2$. For transmitter cooperation, the cut-set capacity upper bound
is found to be
\begin{align}
  \label{eq:Ct2}
  \Ct{2} = \Bigl\{\textstyle 
    \C\bigl(\frac{2g}{g+1}\bigr) \text{ if } g\geq1,
    \quad \C\bigl(\frac{1+g}{2}\bigr) \text{ if } g<1 \Bigr.,
\end{align}
with $\rho^*=(g-1)/(g+1)$ if $g\geq1$, and $\rho^*=0$ otherwise.
Incidentally, the bound $\Ct{2}$ in (\ref{eq:Ct2}) coincides with the
transmitter cooperation rate $\Rt{1}$ in (\ref{eq:Rt1}) obtained in
Case~1 for $g\geq1$. Next, the decode-and-forward transmitter
cooperation rate is given by
\begin{align}
  \label{eq:Rt2}
  \Rt{2} = \Bigl\{\textstyle 
  \C\bigl(\frac{2(g-1)}{g}\bigr) \text{ if } g\geq2,
  \quad \C\bigl(\frac{g}{2}\bigr) \text{ if } g<2 \Bigr.,
\end{align}
with $\rho^*=(g-2)/g$ if $g\geq2$, and $\rho^*=0$ otherwise. Similar
to Case~1, the transmitter cooperation rate $\Rt{2}$ in (\ref{eq:Rt2})
is close to its upper bound $\Ct{2}$ in (\ref{eq:Ct2}) when $g\gg1$.

For receiver cooperation, the corresponding cut-set bound resolves to
\begin{align}
  \label{eq:Cr2}
  \Cr{2} = \Bigl\{\textstyle 
  \C(1) \text{ if } g\geq1,
  \quad \C\Bigl(\frac{1+\sqrt{g(2-g)}}{2}\Bigr) \text{ if } g<1
  \Bigr., 
\end{align}
with $\rho^* = 0$ for $g\geq1$, and $\rho^*=(\sqrt{2-g}-\sqrt{g})/2$
otherwise. Lastly, the compress-and-forward receiver cooperation rate
is 
\begin{align}
  \label{eq:Rr2}
  \Rr{2} = \C\bigl(\textstyle\frac{g}{2(g+2+2/P)}+\frac{1}{2}\bigr).
\end{align}

\begin{figure}
  \centering
  \includegraphics[width=8cm]{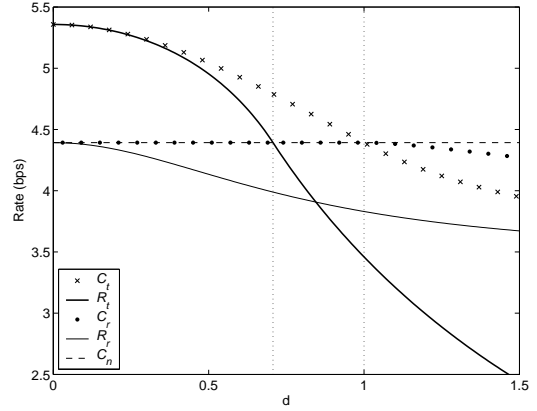}
  \caption{Cut-set bounds and achievable rates in Case~2.}
  \label{fig:rates_2}
\end{figure}

It can be observed that if the cooperating nodes are close together
such that $g>2$, the transmitter cooperation rate $\Rt{2}$ is strictly
higher than the receiver cooperation cut-set bound $\Cr{2}$;
therefore, transmitter cooperation conclusively outperforms receiver
cooperation when the system is under equal power allocation with full
CSI\@.  Fig.~\ref{fig:rates_2} illustrates the transmitter and
receiver cooperation upper bounds and achievable rates.

\subsection*{Case 3: Optimal power allocation with receiver phase CSI}
\label{sec:case3}

When remote phase information is not available, it was derived in
\cite{kramer04:coop_cap_relay, host-madsen04:power_relay} that it is
optimal to set $\rho=0$ in the cut-set bounds (\ref{eq:Ct}),
(\ref{eq:Cr}), and the decode-and-forward transmitter cooperation rate
(\ref{eq:Rt}). Intuitively, with only receiver phase CSI, the relay
and the transmitter, being unable to realize the gain from coherent
combining, resort to sending uncorrelated signals.

The receiver cooperation strategy of compress-and-forward, on the
other hand, did not make use of remote phase information
\cite{kramer04:coop_cap_relay}, and so the receiver cooperation rate
is still given by (\ref{eq:Rr}) with the power allocation parameter
$\alpha$ optimally chosen.

Under the transmitter cooperation configuration, the cut-set bound is
found to be
\begin{align}
  \label{eq:Ct3}
  \Ct{3} = \C(1),
\end{align}
where $\alpha^*$ is any value in the range $[1/(g+1), 1]$. When
the relay is close to the transmitter ($g\geq1$), the
decode-and-forward strategy is capacity achieving, as reported in
\cite{kramer04:coop_cap_relay}. Specifically, the transmitter
cooperation rate is given by
\begin{align}
  \label{eq:Rt3}
  \Rt{3} = \Bigl\{\textstyle 
  \C(1) \text{ if } g\geq1,
  \quad \C(g) \text{ if } g<1 \Bigr.,
\end{align}
where $\alpha^*$ is any value in the range $[1/g, 1]$ if $g\geq1$, and
$\alpha^*=1$ otherwise.

For receiver cooperation, the cut-set bound is
\begin{align}
  \label{eq:Cr3}
  \Cr{3} = \Bigl\{\textstyle 
  \C\bigl(\frac{2g}{g+1}\bigr) \text{ if } g\geq1,
  \quad \C(1) \text{ if } g<1 \Bigr.,
\end{align}
where $\alpha^*=g/(g+1)$ if $g>1$, $\alpha^*=1$ if $g<1$, and
$\alpha^*$ is any value in the range $[g/(g+1), 1]$ if $g=1$. Since
compress-and-forward does not require remote phase information, the
receiver cooperation rate is the same as (\ref{eq:Rr1}) given in
Case~1: $\Rr{3}=\Rr{1}$. Note that the argument inside $\C(\cdot)$ in
(\ref{eq:Rr1}) is 1 when $\alpha=1$, and hence $\Rr{3}\geq\C(1)$.

\begin{figure}
  \centering
  \includegraphics[width=8cm]{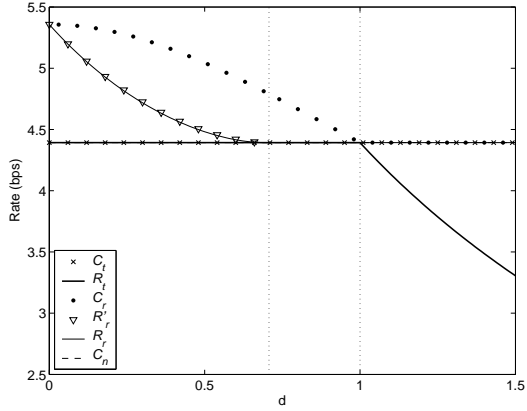}
  \caption{Cut-set bounds and achievable rates in Case~3.}
  \label{fig:rates_3}
\end{figure}

In contrast to Case~2, the receiver cooperation rate $\Rr{3}$ in
equals or outperforms the transmitter cooperation cut-set bound
$\Ct{3}$; consequently receiver cooperation is the superior strategy
when the system is under optimal power allocation with only receiver
phase CSI\@. Numerical examples of the upper bounds and achievable
rates are shown in Fig.~\ref{fig:rates_3}.

\subsection*{Case 4: Equal power allocation with receiver phase CSI}
\label{sec:case4}

With equal power allocation, $\alpha$ is set to $1/2$. With only
receiver phase CSI, similar to Case~3, $\rho=0$ is optimal for the
cut-set bounds and decode-and-forward rate. Therefore, in this case no
optimization is necessary, and the bounds and achievable rates can be
readily evaluated.

For transmitter cooperation, the cut-set bound and the
decode-and-forward rate, respectively, are
\begin{align}
  \label{eq:Ct4}
  \Ct{4} & = \Bigl\{\textstyle 
  \C(1) \text{ if } g\geq1,
  \quad \C\bigl(\frac{1+g}{2}\bigr)  \text{ if } g<1 \Bigr.,\\
  \label{eq:Rt4}
  \Rt{4} & = \Bigl\{\textstyle 
  \C(1) \text{ if } g\geq2,
  \quad \C\bigl(\frac{g}{2}\bigr) \text{ if } g<2 \Bigr..
\end{align}
For receiver cooperation, the cut-set bound is
\begin{align}
  \label{eq:Cr4}
  \Cr{4} = \Bigl\{\textstyle
  \C(1) \text{ if } g\geq1,
  \quad\C\bigl(\frac{1+g}{2}\bigr)\text{ if } g<1\Bigr.,
\end{align}
and the compress-and-forward rate is the same as (\ref{eq:Rr2}) in
Case~2: $\Rr{4}=\Rr{2}$. 

\begin{figure}
  \centering
  \includegraphics[width=8cm]{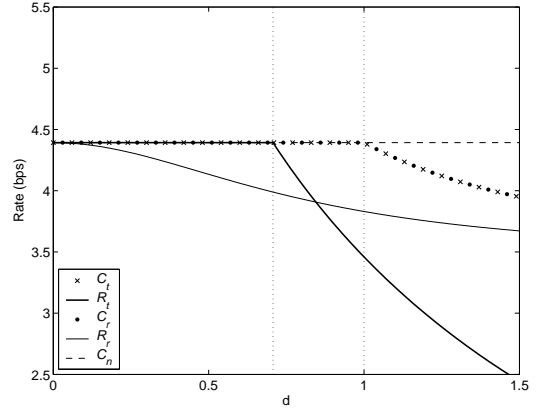}
  \caption{Cut-set bounds and achievable rates in Case~4.}
  \label{fig:rates_4}
\end{figure}

Parallel to Case~1, the transmitter and receiver cooperation cut-set
bounds $\Ct{4}$ and $\Cr{4}$ are identical. Note that the
non-cooperative capacity $\Cn$ meets the cut-set bounds when $g\geq1$,
and even beats the bounds when $g<1$. Hence it can be concluded
cooperation offers no capacity improvement when the system is under
equal power allocation with only receiver phase CSI\@. Numerical
examples are plotted in Fig.~\ref{fig:rates_4}.

\section{Implementation Strategies}
\label{sec:impstrat}

In the previous section, for each given operational environment we
derived the most advantageous cooperation strategy. The available mode
of cooperation is sometimes dictated by practical system constraints,
however. For instance, in a wireless sensor network collecting
measurements for a single remote base station, only transmitter
cooperation is possible. In this section, for a given transmitter or
receiver cluster, the trade-off between cooperation capacity gain and
implementation complexity is investigated.

\begin{figure}
  \centering
  \includegraphics[width=8cm]{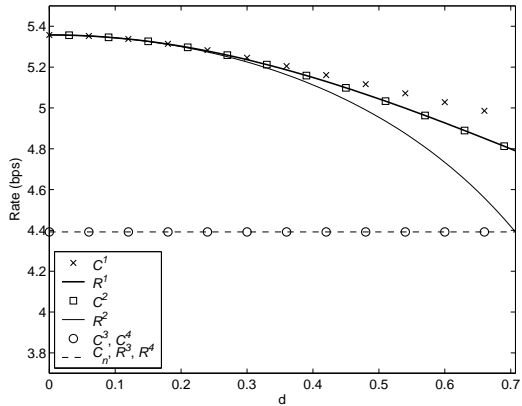}
  \caption{Transmitter Cooperation}
  \label{fig:rates_tx}
\end{figure}

The upper bounds and achievable rates from the previous section are
summarized, and ordered, in Table~\ref{tab:coop_order}: the rate of an
upper row is at least as high as that of a lower one. It is assumed
that the cooperating nodes are close together such that $g>2$. The
transmitter cooperation rates are plotted in Fig.~\ref{fig:rates_tx}.
It can be observed that optimal power allocation contributes only
marginal additional capacity gain over equal power allocation, while
having full CSI is essential to achieving any cooperative capacity
gain. Accordingly, in transmitter cooperation, homogeneous nodes with
common battery and amplifier specifications can be employed to
simplify network deployment, but synchronous-carrier should be
considered necessary.

\begin{table}
  \renewcommand{\arraystretch}{1.6}
  \centering
  \begin{tabular}{|l|l|}
    \hline
    \emph{Cooperation Scheme} & \emph{Rate}\\
    \hline\hline
    \Ct{1}, \Cr{1} & $\C\ns\left(\frac{2(g+1)}{g+2}\right)$\\
    \hline
    \Rt{1}, \Ct{2}, \Cr{3} & $\C\ns\left(\frac{2g}{g+1}\right)$\\
    \hline
    \Rt{2} & $\C\ns\left(\frac{2(g-1)}{g}\right)$\\
    \hline
    \Rpr\ & $\C\ns\left(\frac{2g(\sqrt{g-1}-1)(g-1-\sqrt{g-1})}
      {\sqrt{g-1}(g-2)^2}\right)$\\
    \hline
    \Rr{1}, \Rr{3} & ${\displaystyle\max_{0\leq\alpha\leq1}}
    \C\ns\left(\frac{\alpha(1-\alpha)g}{(1-\alpha)g+2\alpha+1/P}
      +\alpha\right)$\\
    \hline
    \Cn, \Cr{2}, \Ct{3}, \Rt{3}, \Ct{4}, \Rt{4}, \Cr{4} & $\C(1)$\\
    \hline
    \Rr{2}, \Rr{4} & $\C\ns\left(\frac{g}{2(g+2+2/P)}+\frac{1}{2}\right)$\\
    \hline
  \end{tabular}
  \caption{Cooperation rates comparison}
  \label{tab:coop_order}
\end{table}

\begin{figure}
  \centering
  \includegraphics[width=8cm]{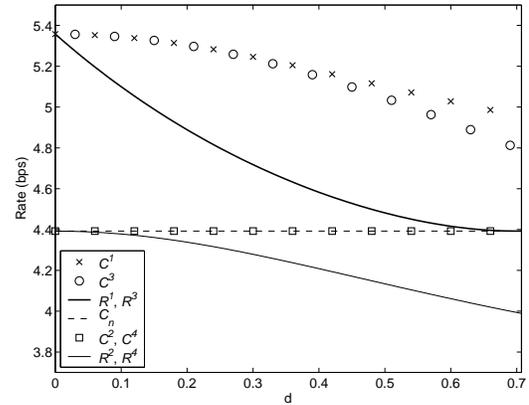}
  \caption{Receiver Cooperation}
  \label{fig:rates_rx}
\end{figure}

On the other hand, in receiver cooperation, the compress-and-forward
scheme does not require full CSI, but optimal power allocation is
crucial in attaining cooperative capacity gain, as illustrated in
Fig.~\ref{fig:rates_rx}. When remote phase information is not utilized
(i.e., $\rho=0$), as noted in \cite{host-madsen04:coop_bounds},
carrier-level synchronization is not required between the relay and
the transmitter; implementation complexity is thus significantly
reduced. It is important, however, to allow for the network nodes to
have different power requirements and power allocation be optimized
among them.

\section{Conclusion}

We have studied the capacity improvement from transmitter and receiver
cooperation when the cooperating nodes form a cluster in a relay
network. It was shown that electing the proper cooperation strategy
based on the operational environment is a key factor in realizing the
benefits of cooperation in an ad-hoc wireless network. When full CSI
is available, transmitter cooperation is the preferable strategy.  On
the other hand, when remote phase information is not available but
power can be optimally allocated, the superior strategy is receiver
cooperation. Finally, when the system is under equal power allocation
with receiver phase CSI only, cooperation offers no capacity
improvement over a non-cooperative single-transmitter single-receiver
channel under the same average network power constraint.

\bibliographystyle{IEEEtran.bst}
\bibliography{IEEEabrv,coop}

\end{document}